# LoRaWAN Temperature Sensors for Local Government Asset Management

ICP Innovation Internship

Jack Downes
Summer Intern
Innovation Central
Perth, Australia
jack.downes@student.curtin.edu.au

*Abstract*—The purpose of this project is to investigate the suitability of using LoRaWAN technology to conduct temperature studies on local government assets in Australian metropolitan and residential areas. Temperature sensing devices were integrated into the existing LoRaWAN infrastructure at Curtin University with data collected and stored on a remote server. Case studies were performed for the City of Melville to address the suitability for such a system to provide insights into heat islands and urban forests. Testing was completed on the Curtin University campus, replicating the climate conditions, asset types and, dense building and tree environment found in the City of Melville.

## I. INTRODUCTION

### A. Motivation

The City of Melville is a local government entity in Perth, Western Australia. Covering a metropolitan area to the South West of Perth's CBD the City of Melville is home to 98 083 residents [1]. The City of Melville has been working with the Curtin University Spatial Sciences Department to decipher spatial information related to heat dissipation in their suburbs.

Heat Islands are areas related spatially by their heat retention properties. This often occurs as a result of the use of man-made materials when constructing environments. The phenomena can be observed in Perth's hot months during the night, as the ambient heat in areas using dense materials is noticeably higher than surrounding areas. Materials responsible for creating heat islands include concrete, asphalt and paving. This is an important issue for residence due to the unpleasantness of the heat, and financial pressure caused by the cost of air conditioning.

Urban Forests are areas of vegetation found in metropolitan environments. Through shading of the ground and absorption of thermal energy these areas have a cooling effect on ambient air temperatures. Overnight in hot Perth months these areas will have a comparatively lower temperature than the Heat Islands mentioned above.

It is in the interest of the City of Melville to mitigate the effects of Heat Islands and fully utilise the cooling effect of Urban Forests. Getting this balance right poses a number of questions that can affect budget, maintenance requirements and the livability index of an area. Thermal imagery has been used to assess the effects of Heat Islands and Urban Forests. Thermal imagery can be captured aerially or terrestrially providing a radiation map that can be used to get an understanding of a surface's heat properties.

The City of Melville in conjunction with Curtin University's Department of Spatial Sciences conducted three studies in 2017 to address interests in Heat Islands and Urban Forests. Bradley Schupp's paper [2] analyses the City of Melville's aerial thermal imagery and LIDAR data investigating land use and Heat Island correlation. Stuart McEvoy's paper [3] looks into cooling effects by type of vegetation using terrestrial thermal imagery and on foot temperature sampling. Luke Ellison's paper [4] links the changing of land use types to the prominence of Heat Islands.

The nature of this research lends itself to bigger datasets and higher granularity of data.

### B. Scope

To ensure the viability of this internship project, it is important to limit the project scope. This is defined as:
- Implement a LoRaWAN system to wirelessly record ambient temperature data
- Evaluate the effectiveness of this use case for the City of Melville.

### C. Thesis Statement

Investigate the suitability of using LoRaWAN technology to conduct research on City of Melville assets in the form of a short term deployments of an array of temperature sensor devices.

## II. BACKGROUND

LoRaWAN is a trending solution for implementation of the Internet of Things style wireless devices. The following helps highlight its value, providing depth of knowledge as to why it may be of benefit to the City of Melville for temperature studies of in future.

### A. The Internet of Things

The Internet of Things (IOT) is a concept born out of the evolution of the Internet. With the increasing reach of networks and the low cost of transmitting data it is possible to connect more devices to the internet. So much so that objects often thought of as mundane and serving a single purpose can be connected and addressable through internet enabled devices to open up a realm of informational possibility. This allows for network interaction with existing everyday items such as home appliances, building climate control, street lights, or car parks. Through this trend large datasets of information can be gathered and used to improve processes conventionally ignored. The concepts of IOT are largely accepted as the direction of the Internet of the future [5]. As enabling technologies reach a level of maturity and standards are adopted it is expected that IOT will be adopted as a mainstay of smart cities [6].

### B. LoRaWAN

Low-Power, Wide-Area Networks (LPWAN) are networks with a priority on keeping power usage for transmitting data to a minimum. They are designed specifically for battery powered devices, having the potential to extend expected battery life cycles to years. This battery performance is not possible through common wireless networks like WI-FI. LPWANs will support a large portion of the billions of devices forecasted for the IOT [7].

LoRaWAN (LoRa Wide-Area Network) is a realisation of the LPWAN concept and is becoming a popular solution [8].

LoRaWAN is a standardised system architecture that implements LoRa, a low power long range communication protocol using technology not available for commercial use prior to its release in June 2016 [7] [9]. To achieve low power transmission the data throughput is restricted, meaning it is unsuitable for transmission of large data necessary to transmit video or audio. This system is designed to be a solution for small packets sent at large intervals, such as weather station data, non-frequent GPS locations or gas sensor readings.

Figure 1 shows the LoRaWAN structure. The system is composed of four parts; the end nodes, LoRaWAN gateways, a network server and application servers. Together these components act to read sensor data, transmit it wirelessly to a gateway, and repeat it through the internet onto an application server to be accessed by an end user.

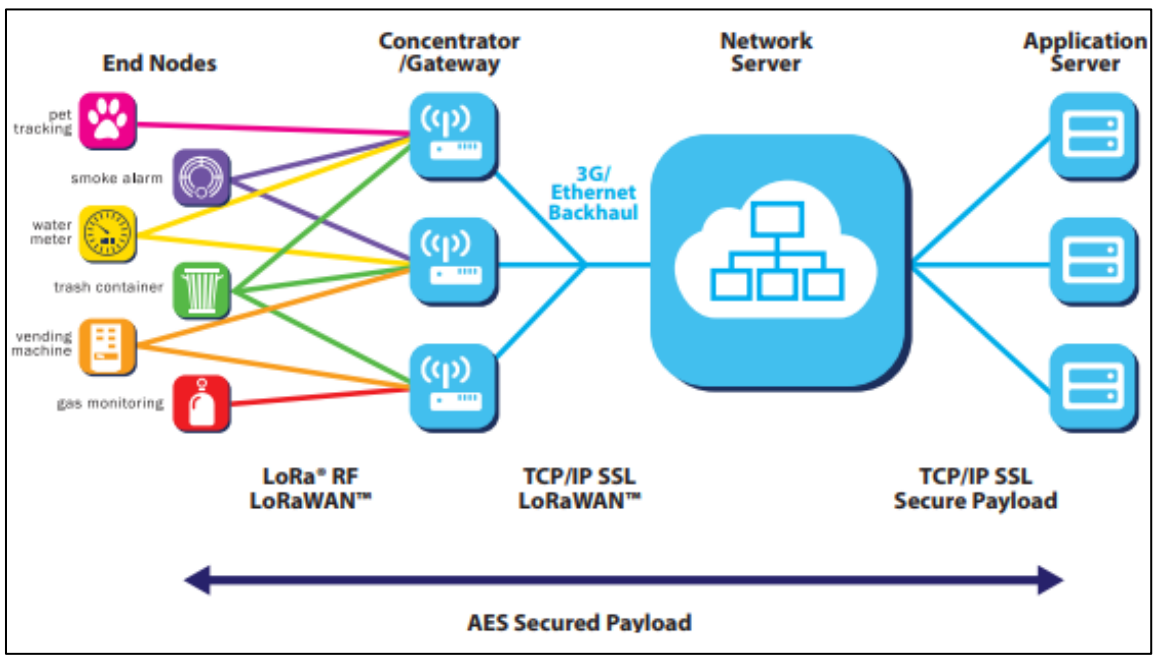

*Figure 1 - LoRaWAN System Architecture [7]*

### III. IMPLEMENTATION

A LoRaWAN system for the City of Melville was built. It would allow insight into the practicality of the proposed use cases. The new method of data capture would then be available for trials replicating the findings of the previous studies. A break-down of the system design follows.

## A. End Node Device

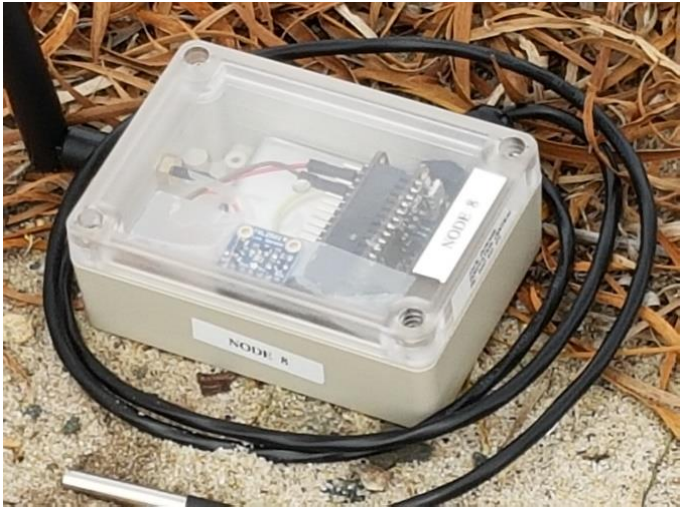

*Figure 2 - Temperature Sensor Device*

Three sensor devices were created to form the array of end nodes. The array allows the compilation of data from many devices in unison. Using the Cisco Wireless Gateway [10] the size of the array can be expanded far beyond these three devices. Having three sensors allows comparison of asset material properties that are exposed to simultaneous variables caused by weather.

Each temperature sensor device (shown in Figure 2) was built to collect ambient air temperature and light intensity, transmitting it through to the common parts of a LoRaWAN. The device contains an Adafruit Feather M0 RFM95 LoRa Radio microcontroller, rechargeable 3.7V 2500mAh battery, Waterproof DS12B20 digital temperature sensor (on a 900mm long cable), TSL2561 digital luminosity/LUX/light sensor, and a UFL antenna. It is contained within a sealed plastic enclosure, providing protection against dust and low pressure water jets [11].

The external waterproof temperature sensor and internal light sensor provide complementary data. The temperature sensor records ambient air readings above a surface of interest which in daylight hours is affected by direct sunlight. By recording light intensity at the same time, direct sunlight could be identified as a contributing factor to a high temperature reading. In turn this extra information helping to rule out outliers in data sets.

## B. Gateway and Network Server

The gateways and network server were implemented using existing infrastructure available to Innovation Central Perth. Two gateway antennae exist at Curtin University's Bentley Campus. One above building 216, and another atop building 105 (the Robertson Library). The network server used for the system was hosted on Cisco infrastructure. When receiving a packet of information from a sensor device the gateway forwards it onto the network server. A web hook hosted on the network server was programmed to interrogate packets passing through, identify those belonging to this project, and send the appropriate information to the application server through a MYSQL connection.

## C. Application Server

The Database is hosted within Cisco's Virtual Private Network (VPN) on a virtual machine (VM) running Ubuntu Server 16.04. The server contains the structural ability to store information about each device unit, projects, user details, material properties, location properties as well as the readings themselves. It was important to structure the data in this way to retain its integrity and accessibility. Even at a low number of nodes data management is an important aspect to the success of a LoRaWAN system. To access data from the server the database must be queried using a frontend program, e.g. MYSQL Workbench, then the requested data is saved to a client PC to be filtered and analysed.

## IV. RESULTS

Using a deployment plan to manage the various case studies in parallel the devices could be loaded with the properties required for each assignment, taken to the location of interest and deployed. Data was captured and stored for post analysis, as live feed data was not considered important for this use case.

The datasets produced rapidly exceeded readings in excess of ten thousand timestamped entries. As such, tools like Microsoft Excel were not suited to interrogating the size of the dataset. This led to the use of the data analytics tools available through Python packages. Python's Spyder IDE along with the pandas, numpy and matplotlib libraries allowed filtering of data for gross errors, periods of interest and trends, and the ability to plot the data in a digestible format. The plots following and found in the appendix were generated using this process.

The plots proved to be of good use for quantitative analysis. A case study into the different cooling effects of concrete and grass measured the dissipation of heat of the surface coverings in parallel over the same night. This study replicates the interests the City of Melville has in Heat Islands and Urban Forests. Figure 6 outlines the expected behaviour. After moving into the shade of the day, concrete begins to dissipate heat, but at a slow rate. The ambient temperature above the concrete by the time it begins to reheat from the sun sits just above 20°C. Conversely by 18:30 the grass rapidly drops to below 20°C within 4 hours. This evidence of materials dissipating heat at different rates can be easily replicated, showing the value in having a quick way of attaining large data sets.

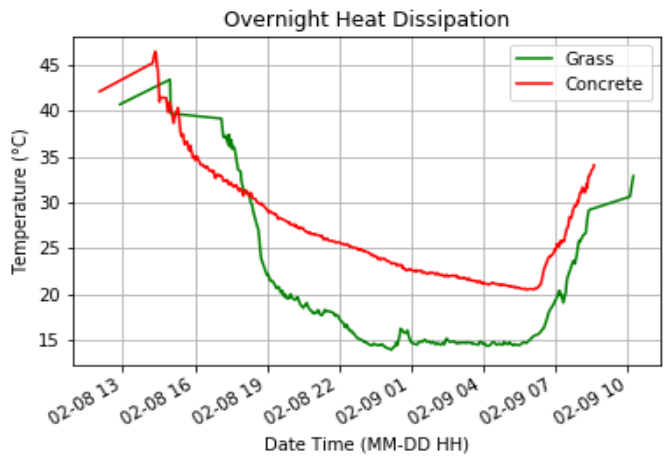

Figure 3 - Heat dissipation overnight with grass and concrete

Data from the system also lends itself to comparison with third party data sets. Figure 8 shows temperature fluctuation from a node's sensor that was implemented on a red brick wall in an area receiving sunlight over 4 days, plotted against dry bulb temperatures sourced from a weather station run by the Bureau of Meteorology at the Perth Airport [12] ten kilometres away from the Curtin University's Bentley Campus. It outlines the behaviour of material temperatures when the effects of sunlight are no longer present. Any timestamped data can be used in conjunction with data captured through the LoRaWAN system.

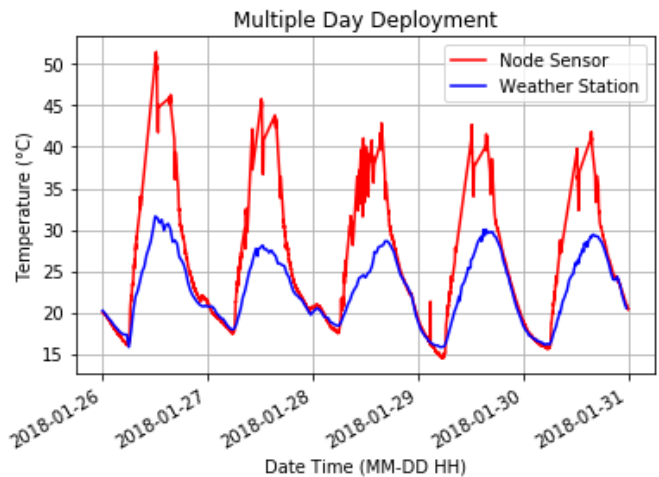

Figure 4 - Heat dissipation over multiple days with weather station data and a node setup on red brick

## V. CHALLENGES

Building a LoRaWAN system with the help of Innovation Central Perth proved to be straight forward. The hardware and network components parts of the project came together without fault. This allowed the use case specific challenges to become abundantly clear. They are as follows:

A. *Network Coverage*

LoRaWAN can achieve transmission distances of five kilometres with line of sight between the gateway and end node antennae, however poor network reception was encountered. This was due to non-terrain obstructions like trees and buildings. These network issues became apparent five hundred and fifty metres away from the gateway, when the antenna was low to the ground. This can be mitigated by altering the end node design to allow for raising the antennae from the ground, installing more gateways in the area of interest to create an overlapping coverage, or utilising temporary gateways on location connected through a mobile network.

B. *Overheating Nodes and Direct Sunlight*

At surface temperatures of around 40℃ the microprocessor within the end node device turns off, avoiding damage caused by the electronics operating at high temperatures. Figure 5 shows two full days of deployment for a device recording red brick, the time scale annotation shows the month (January), the day (the 26th to the 27th) and the hour in 24-hour time. It can be observed here how direct sun light affects devices readings in two ways. Firstly the deviation of the surface temperature from the dry bulb temperature readings. Secondly, periods of no readings, or gross errors experienced during the heat of the day between 10:00 and 16:00 hours on both the 26th and the 27th of January. To mitigate the second issue for the case studies the device body was kept out of direct sunlight by trailing the external sensor to the area of interest. Through installing a fan, a heat sink, using more appropriate enclosure materials, or using an opaque enclosure lid, the design can be modified to operate in higher temperatures.

In practice the light sensor provided little use, often recording gross errors as a result of the temperatures reached inside the box. It remains a contender as a solution to recording the effects of shade, cloud cover, and the passing of direct sunlight at sunrise and sunset, but its data was not as telling as that collected from the temperature sensor and contained far more noise that needed to be filtered out.

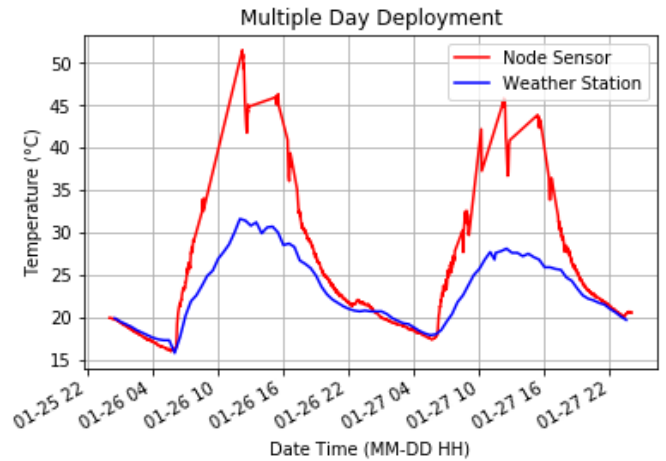

Figure 5 – End node deployed on red brick at Curtin University plotted against Perth Airport weather station data provided by the Bureau of Meteorology [12]

*C. Theft*

With dimensions similar to a modern smart phone, the unit's small size does lend it to the possibility of theft. For this project the devices were most often deployed in areas of low pedestrian traffic that were locked outside of business hours. For the use of local government this will be a rare luxury. In future iterations the design could be altered to mitigate the risk of theft by using a more robust enclosure with the capacity to be locked down, GPS tracking, camouflaging, larger more cumbersome boxes, smaller more discrete boxes or an opaque top.

In one case study a device was deployed in a public area outside the Curtin University Stadium. It was left on an open grass field with a medium level of foot traffic. During this deployment it became evident how discrete the unit is at its small size: it was difficult to see until standing within metres of the unit. As the unit cost for these devices can be quite low, and if theft only happens rarely, allowing for occasional theft may be within reason.

The system is designed to transmit information and store it remotely for every reading, not saving anything on the device. This abstraction means that the impact made by a device being stolen or broken on a project's data set, is far less than for a device storing data within its local storage.

*D. Maintenance*

The system requires maintenance after the initial setup. Some maintenance requirements that have been identified are:

- The battery for each unit needed to be recharged after around two hundred hours of use, or before each deployment (taking around four hours).
- Time needed to be taken to support unexpected software faults.
- The external temperature sensor may rust and batteries will degrade over time, needing to be replaced.

## VI. DISCUSSION

The range of applications for network technology means the LoRaWAN system designed for this project can only address a small use case out of the vast possibilities. Following are points of discussion into how the system may be better utilised for the City of Melville in the future.

*A. User Interface*

Beside the initial setup the system requires an understanding that may take up to a week for the non-inducted user to become familiar with. This includes an understanding of **MYSQL** to pull information from the database, **low level coding** to alter the properties for each sensor node (such as transmission interval) and **data analysis** to interrogate the data. To manage this project the software packages Arduino IDE, MYSQL Workbench and Spyder were used. If all of this knowledge can be encapsulated to one user interface abstracting the low level information from the user, it would make the system much more user friendly. Options include a custom desktop client, a browser client or existing third party software like AdafruitIO [13].

*B. Device Sensor Options*

Once the supporting structure of a LoRaWAN is established, the device array it supports can be developed and improved to better suit the needs of local government. For example, an observation from the data in Figure 5 is the deviation between the dry bulb temperature reading of ambient air and that recorded on the ground in direct sunlight by the device. The effects from the direct sunlight on the external temperature probe could be mitigated using a temperature sensor that relies on radiation emitted from surfaces instead of the electronic resistance based sensor that was implemented, as this may be affected by its own material properties. A study is possible with the system as it stands to add a range of different types of temperature sensors to each node and to investigate how they behave in comparison. The end node devices can be further customised to monitor asset characteristics including soil moisture levels, path and road utilization, air quality, and weather data, along with array of Smart City [14] opportunities.

*C. Set and Forget*

A major benefit of LoRaWAN is its support for set and forget style devices. Set and forget devices are left in the field and do not require any servicing, either achieved by a solar power source or by a battery that services the life of the project, often expected to be years. LoRaWAN is built to enable devices to operate on a single battery charge for more than a year; use cases for local government may call for this capability. Longer deployment periods can allow deeper insights than those achieved in testing for this project. For example, temperature data tracking the growth of forest canopies into maturity, seasonal effects on material heat dissipation, temperature properties relating to long term material degradation, and more accessible real time and retrospective interrogation for events such as heat waves and freak storms.

This project focussed on shorter period deployments to suit the timeframe of the internship. It allowed for shorter transmission intervals of one to five minutes, chosen to achieve a higher temporal granularity for plotting temperature and to account for redundancy in the event of lost data. This higher transmission interval was achieved at the cost of battery life. With a transmission interval of two minutes the end nodes last up to two hundred hours. This can be extended by increasing the recording interval to thirty minutes. Furthermore by adding a low power timer component [15] (a quick task) the devices battery life can be greatly expanded. This component helps to regulate power draw by reconnecting the battery supply at a timed interval, allowing the microcontroller to boot up for only the reading and transmission cycle then turn itself off. This addition can increase the battery utilisation and operating time to better suit set and forget deployment.

## VII. RECOMMENDATIONS

A major barrier to entry for end users of this system is the base knowledge required to navigate the data capture pipeline. To ensure ease of use for the system and a high level of uptake for projects it is recommended to develop a user friendly digital interface. Most important would be the abstraction of downloading data from the application server. Once in a familiar format on a client PC the data can be intuitively analysed. The data analysis process can be stream-lined to output comparative plots and trend information, which are easily digestible for local government use cases. Failing the above, it would be advised to package this solution as a service, to ensure customer expectations are met, and to remove any need for training that may be required.

Network coverage proved a hindrance to the effectiveness of available device locations. This meant some surfaces could not be tested as they were at the fringe of the network range on the Curtin University campus, or had poor line of sight to the gateway antennae. A portable LoRaWAN gateway may provide a solution to this issue by filling in shadows created by obstructive trees or buildings. The time spent trouble shooting packet drop and consequently reassessing the deployment plan in the case study phase of the project highlights network coverage as a key component to the success of using LoRaWAN devices for temperature sensing in the City of Melville. Investigating solutions to improve network coverage in obstructed terrain is recommended.

LoRaWAN has been developed to support devices deployed for long periods of time. It is recommended to investigate extending the battery life of device nodes and compare to that achieved for this project. This will allow use cases for set and forget devices for the City of Melville to be trialled.

## VIII. CONCLUSION

As metropolitan living becomes denser the management of local government assets becomes more complex. As we move further into the information age, there is an opportunity to take advantage of big data solutions like the Internet of Things. The Australian Government's Smart Cities initiative [14] is one example of public effort to realise the transformative potential such technologies. LoRaWAN and sensors, such as those developed in this project, provide a standardised, widely adopted and supported way to capture data to inform Smart Cities now and in the future.

## IX. ACKNOWLEDGEMENT

I'd like to note my sincere gratitude to Innovation Central for the opportunity to work on this project, entrusting me with the room to manage it as I saw fit, to Valerie Maxville for the guidance she provided to myself and my peers, and to Jim Wyatt for encouraging us to take pride in our work. My thanks to Janine Ahola from the City of Melville for fostering excitement for any outcomes the project might produce. And my thanks to Dave Belton and Petra Helmholz from Curtin University's Department of Spatial Science for their professional advice and vision for this project.

# XI. APPENDIX

## A. CASE STUDY 1: Concrete vs Grass

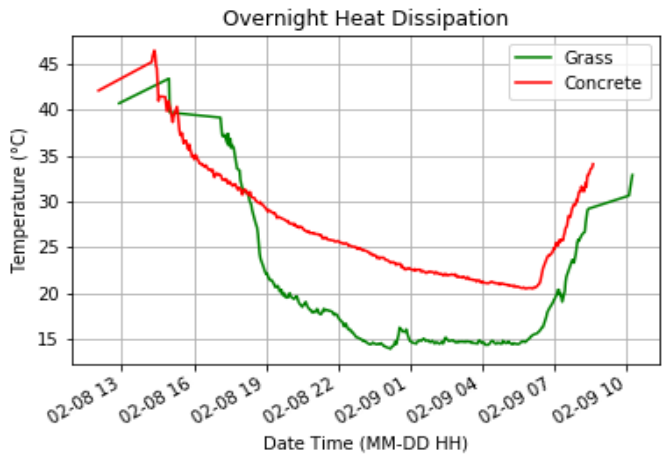

*Figure 6 - Heat Dissipation Overnight Grass and Concrete*

## B. CASE STUDY 2: Playground Materials

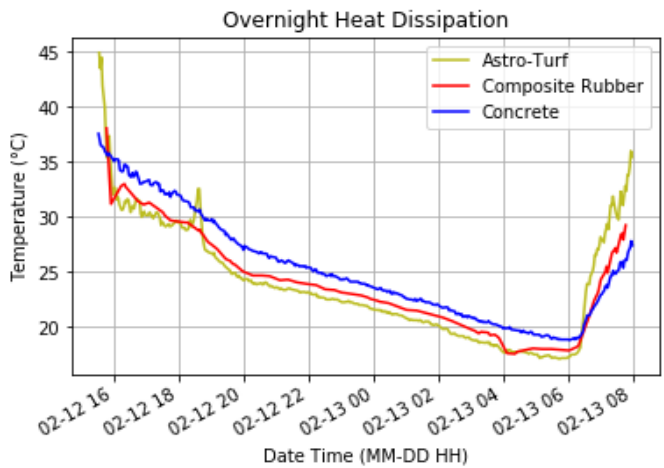

*Figure 7 - Heat Dissipation Overnight Playground Materials*

## C. CASE STUDY 3: Weather Station Integration over a Week

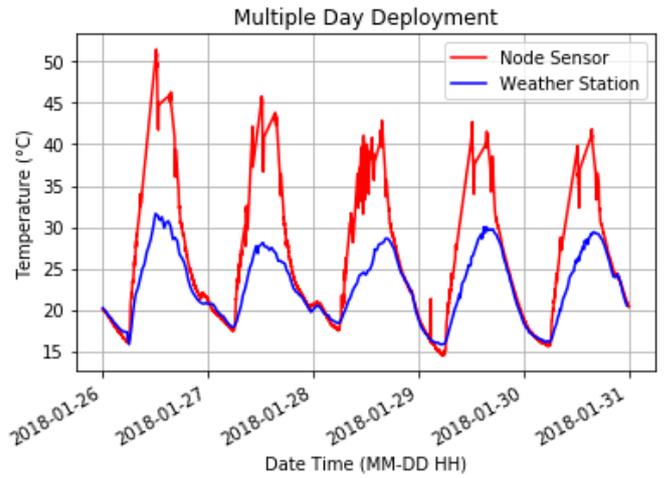

*Figure 8 - - Heat dissipation over multiple days with weather station data and a node setup on red brick*

## D. CASE STUDY 4: Tin and Concrete

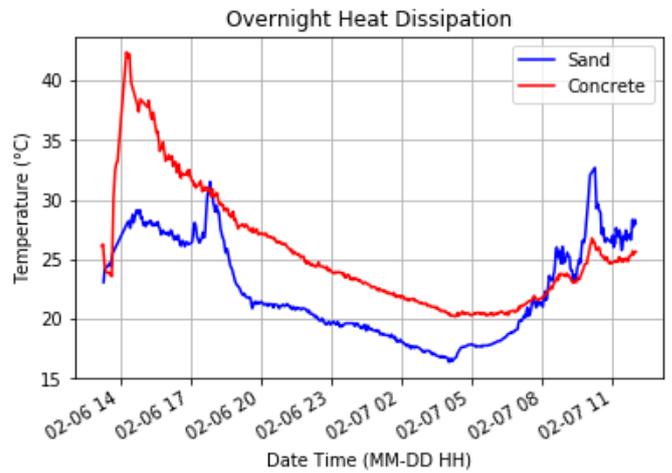

*Figure 9 – Heat dissipation of tin and concrete*